\documentclass{article} \usepackage{epsf} \usepackage{amsmath}
\usepackage{amssymb}
 
\newcommand{\be}{\begin{equation}} 
\newcommand{\ee}{\end{equation}}

\newcommand{\half}{\frac{1}{2}}

\newcommand{\psibar}{\overline{\psi}}

\newcommand{\chibar}{\overline{\chi}}
\newcommand{\Tr}{{\rm Tr}}
\newcommand{\tr}{{\rm tr}}
\newcommand{\Real}{{\rm Re}}

\newcommand{\bsplit}{\begin{equation} \begin{split}}
    \newcommand{\esplit}{\end{split} \end{equation}}

\begin{document}
\thispagestyle{empty} \parskip=12pt \raggedbottom
 
\def\mytoday#1{{ } \ifcase\month \or January\or February\or March\or
  April\or May\or June\or July\or August\or September\or October\or
  November\or December\fi
  \space \number\year}
\noindent
\vspace*{1cm}
\begin{center}
  {\LARGE Chiral symmetry on the lattice}
\footnote{Solicited contribution to the volume 
{\it 50 Years of Yang-Mills Theory}, edited by G.~'t Hooft (World
  Scientific)} 

 
  \vspace{1cm} P. Hasenfratz\\
  Institute for Theoretical Physics \\
  University of Bern \\
  Sidlerstrasse 5, CH-3012 Bern, Switzerland
  
  \vspace{0.5cm}
  
  \nopagebreak[4]
 
\begin{abstract}
As a non-perturbative and  gauge invariant regularization the lattice provides
a tool for deeper understanding of the celebrated Yang-Mills theory, QCD and
chiral gauge theories. For illustration, I discuss some analytic
developments on the lattice related to chiral symmetry,
chiral fermions and improvement programs.
Chiral symmetry on the lattice has an amazing history, and it might influence
our perception of a symmetry beyond this example.
\end{abstract}
 
\end{center}
\eject

\section{Introduction}
After the introduction of non-Abelian gauge theories~\cite{YM} it took a long
time to understand how to set up perturbation theory~\cite{FP} and control the
ultraviolet fluctuations~\cite{HV} in such models. Asymptotic
freedom~\cite{PGW}
raised the possibility to connect non-Abelian gauge theories with the
phenomenology of strong interactions. It was not
clear, however, what to do with the wild infrared divergences present in
perturbation theory. A few courageous theorists with good intuition
interpreted 
this problem as a virtue: objects with color are confined by strong forces 
and the infrared problem in perturbation theory is
an indication of that. Quantum chromodynamics (QCD), the theory of
quarks in interaction with 
gluons in an $SU(3)$ gauge theory became an attractive candidate to
describe the physics of hadrons~\cite{qcd}.

In 1974 Wilson~\cite{Wilson1974} presented a $U(1)$ gauge theory with
fermions on a hyper-cubic four dimensional Euclidean lattice 
which has confinement
in the $g \rightarrow \infty$ limit, where $g$ is the bare gauge
coupling. Although in this limit the model lives on a coarse lattice far from
the continuum this work introduced many of the ideas, tools and notions which
lead to a new field in particle physics. The formulation is fully gauge
invariant and independent of perturbation theory. It creates also a strong
link to 
critical phenomena in statistical physics. The construction was
generalized to $SU(N)$ in the same paper.
The confinement problem can
be formulated in this language the following way: do these models stay in the
confining phase as the bare gauge coupling is tuned towards $g=0$ where the
continuum theory is defined? It took a few years until the new tools (mean
field and strong coupling expansion techniques~\cite{BDI74_75}, Hamilton
formulation~\cite{KS1975}, Monte Carlo simulations~\cite{CJR1979}) have been
adapted to quantum field theories, in particular to QCD.

In 1980 Creutz demonstrated~\cite{Creutz1980} in a Monte Carlo simulation that
close to the continuum limit the string tension $\sigma$
between a static quark-antiquark pair in an $SU(2)$  
Yang-Mills theory behaves as expected from continuum renormalization group
considerations. This was a spectacular result indicating that one can get
close to the continuum with modest computing power in this non-Abelian gauge
theory and confinement survives this limit. Of course, 
today's state-of-the-art calculations control 
the systematic and statistical errors of the static
quark-antiquark potential both in $SU(2)$ and $SU(3)$ gauge theories much
better. The potential is followed up to $4\,$fm distances, 
the fine details of the
fluctuating color flux tube are seen and quantitatively
determined~\cite{tube2004}. Although it is not proven rigorously,
there remains little doubt that in these gauge theories
the static quarks are confined in the continuum limit.

QCD, the theory of strong interactions is, however, a Yang-Mills gauge theory
coupled to dynamical quarks. It has an extremely rich phenomenology most of
which is beyond the reach of perturbation theory. QCD is an exciting theory in
itself. In addition it enters most of the processes in the standard model in a
relevant way either in hadronic matrix elements, or as a background. For a
recent general overview on the theoretical and numerical activities in lattice
QCD I refer to ref.~\cite{DeG} and the references therein.

In this contribution I will mainly consider the history of chiral symmetry on
the lattice which was long and troubled and had an amazingly nice upshot. This
outcome might influence our perception on the realization of a symmetry in a
quantum field theory in general. The following overview is  
non-technical. For further details and applications I refer to the 
summaries in \cite{feri,summaries,giulu}.   

\section{QCD on the lattice}
A hyper-cubic lattice is a natural regularization when
path integrals are used to formulate a quantum field theory in $d=4$ Euclidean
space. The quark field $\psi$ lives on the
lattice points indexed by an integer vector $x\in Z_4$ while the gauge fields
are associated with the links $(x,{\hat\mu}), \mu=1,\dots,4$ of the
lattice. Derivatives are replaced by finite difference operators using, for
example the nearest neighbor forward and backward difference
operators\footnote {Here and in most of the following expressions we suppress
the lattice unit $a$. All the fields and parameters are made dimensionless by
appropriate $a$ 
powers: in the mass $m$ a factor of $a$, in the fermion field $\psi$ a factor
of $a^{3/2}$, etc, is absorbed.}
\begin{equation}
\label{nnlat}
\partial^\mu_{x,y}=\delta_{x+\hat{\mu},y}-\delta_{x,y},\qquad
\partial^{\mu*}_{x,y}=\delta_{x,y}-\delta_{x-\hat{\mu},y}\,.
\end{equation}
A simple guess for the action of a massless free fermion on the lattice
might read as 
\begin{equation}
\sum_{x,y} \psibar_x \half \gamma^\mu\left(\partial^\mu_{x,y} +
  \partial^{\mu*}_{x,y}\right)\psi_y \,.
\end{equation}
Switching on the gauge interaction a parallel transporter is included in the
finite difference operators to assure gauge invariance
\begin{equation}
\label{gauder}
\nabla^{\mu}_{x,y}=U(x,\mu)\,\delta_{x+\hat{\mu},y}-
\delta_{x,y}\,,\qquad
\nabla^{\mu*}_{x,y}=\delta_{x,y}-U(x-
\hat{\mu},\mu)^\dagger\,\delta_{x-\hat{\mu},y}\,,  
\end{equation}
where $U(x,\mu)$ is an element of the gauge group $SU(3)$. 
The gauge field $U(x,\mu)$
can be expressed in terms of the vector potential $A_\mu(x)$ used in the
continuum formulation as
\be
\label{transporter}
U(x,\mu) = P\exp[ i \int^1_0 d\tau A^\mu(x+ (1-\tau)\hat{\mu})]\,,
\ee 
where $P$ denotes $\tau$-ordering and $A_\mu$ is an element of the Lie
algebra. On the lattice, however, the group element $U(x,\mu)$ is the
fundamental field variable which transforms under a gauge transformation as
\be
\label{gtrafo}
\tilde{U}(x,\mu)= V(x) U(x,\mu) V^\dagger(x+\hat{\mu})\,.
\ee
Eq.~\eqref{gtrafo} implies that the trace of the product of
gauge matrices along a closed
path on the lattice is a gauge invariant quantity. Locality requires that the
extension of the gauge action is $O(a)$. Taking the smallest loop, 
the plaquette, we get a discretized, local and gauge invariant action for QCD
\be
\label{naiveqcd}
S = -\frac{2}{g^2} \sum_{p} \Real \Tr\, U_p + 
\sum_{x,y} \psibar_x\,D_{x,y}\,\psi_y \,,
\ee
where the sum runs over the plaquettes, $U_p$ is the product of four directed
link matrices around the plaquette $p$ and the Dirac operator has the form
\be
\label{naiveD}
D=\half \gamma^\mu \left(\nabla^\mu +
\nabla^{\mu*}\right) \,.
\ee
Indeed, taking the formal continuum limit $a \rightarrow 0$ of
eq.~\eqref{naiveqcd} and using eqs.~(\ref{gauder},\ref{transporter}) we
obtain the standard continuum form of the QCD action for a massless quark.  
Obviously, there are infinitely many local and gauge invariant lattice actions
which have this property. The continuum predictions must not depend on this
freedom, however. In an asymptotically free theory the scaling dimension of
operators agrees with their engineering dimension up to logarithms (at least
in perturbation theory) and it is easy to count the number of relevant
operators. Perturbative renormalizability can be demonstrated on the lattice
also~\cite{Reisz}.  

In order to define the quantum theory we have to fix the integration measure
of the path integral. Here we take the natural choice of the group invariant
measure for every link variable $U(x,\mu)$ and integrate over the Grassmann
variables $\psibar(x)$ and $\psi(x)$ according to the standard rules of
Grassmann integration. This measure is also gauge invariant.

The action in eq.~\eqref{naiveqcd} has a problem, however.
The Dirac operator in eq.~\eqref{naiveD} describes
actually 16 massless fermion species - a problem called doubling. This is the
first small surprise along a tortured way around chiral symmetry.  

\section{Doublers, chiral symmetry and a \\
no-go theorem}
The free fermion Dirac operator $D$ in Fourier space has the form 
\be
D(p)=\sum_\mu \gamma^\mu \sin(p^\mu)
\ee
which has 16 zeros in the first Brillouin zone at $(0,0,0,0),\dots,
(\pi,\pi,\pi,\pi)$. It can be shown~\cite{KaSm} that these species couple
to the axial current with alternating sign and their total contribution to the
$U(1)$ axial anomaly cancels
~\cite{Adler,BJ} (see also the contributions of Adler and
Jackiw in this volume). This observation resolves a paradox: the
theory is fully regularized, $D$ anticommutes with a standardly
defined $\gamma_5$, so we have a U(1) chiral symmetry without anomaly. Without
the doublers this would be a contradiction.

The question remained whether there exists a clever lattice discretization
such that the doublers are removed but (at least the non-singlet part of)
chiral symmetry survives on the
lattice. This question was answered by a no-go 
theorem~\cite{NN1981} (see also ref.~\cite{KaSm}).
The no-go theorem states that for free fermions the following four conditions 
can not hold
simultaneously:\\
1. $D(x)$ is local\,;\\
2. the Fourier transformed $D$ behaves for $p \ll 1$ as $i\gamma^\mu p^\mu +
   O(p^2)$\,;\\
3. there are no doublers\,; \\
4. $\gamma_5 D + D \gamma_5$ =0\,.

If we do not want to violate flavor symmetry for massless fermions then $D(x)$
is diagonal in the flavor indices and the statement is valid for each flavor
separately.
Locality is the most important property of a quantum field theory, so we can
not give up this condition. The 2nd point just declares that we have a 
Dirac particle. If we do not want doublers (since there are no such copies in
nature), there remains only the possibility to violate condition 4, i.e. to
give up chiral symmetry on the lattice. 

Accepting chiral symmetry breaking terms in the action there are different
possibilities to follow. Wilson~\cite{WilsonErice} suggested to add a
dimension 5 operator to the action in eq.~\eqref{naiveqcd} which modifies 
the Dirac operator as 
\be
\label{Dwilson}
D_{\rm W}=\half \left[\gamma^\mu (\nabla^\mu+\nabla^{\mu*}) 
           -\nabla^\mu \nabla^{\mu*}\right] \,.
\ee
The new term (Wilson term) gives large $\propto$ cutoff masses to the 
doublers, but
- at least classically - leaves the $p=0$ pole unchanged. The Wilson term has
an effect on the renormalization of bare parameters (in particular, for the
bare fermion mass $m_q=0$ is not protected), creates mixings between operators
in different chiral multiplets and it also influences the way and
speed the continuum is approached. The final physical predictions of the
continuum theory are, however, independent of this term.

One might keep a part of the chiral symmetry by giving up flavor symmetry on
the lattice~\cite{KS1975,staggered}. The Kogut-Susskind (or 'staggered') 
fermions
describe 4 flavors and keep a (non-singlet) U(1) part of the original
$U(4)\times U(4)$ symmetry. This formulation has many attractive properties,
in particular for numerical simulations~\cite{qcdsolved}. On the other hand,
getting QCD with 3 light flavors is nontrivial with staggered fermions. I
shall not follow this interesting possibility further in this contribution
since staggered fermions are not directly related to the developments I want
to discuss.

\section{The Ginsparg-Wilson relation}
The no-go theorem seemed to close all the ways towards a chiral symmetric
lattice regularization. 
As the authors formulated~\cite{NN1981}: 'The important
consequence of our work is to {\it discourage any attempt to construct chiral
invariant lattice models for QCD}'. No-go theorems, however, are frequently 
circumvented in an unexpected way. This is what happened with chiral symmetry 
on the lattice.

Soon after the no-go theorem was presented Ginsparg and Wilson~\cite{GW1982}
suggested a condition for the Dirac operator which, as the authors
formulated, implies a
'remnant' chiral symmetry on the lattice. More than 15 years later this
condition, called the Ginsparg-Wilson relation, became very important 
in the theoretical developments concerning global and local chiral symmetry on
the lattice and beyond.
 
Ginsparg and Wilson treated free fermions and used
some basic notions of Wilson's renormalization group theory. 
Consider a renormalization group transformation which blocks the fermion field
$\psi_x$ living on the fine lattice into the $\chi_{x_B}$ fermion
field on the coarse lattice:
\be
\label{GWblock}
\begin{split}
 \exp(-S_B (\chibar,\chi))& = \int D\psibar D\psi \,  
\exp[\,-S(\psibar,\psi) \\
  & - \sum_{x_B,y_B}
 (\chibar_{x_B} - \sum_x \psibar_{x}\,\omega^\dagger_{x,x_B}) \,
R^{-1}_{x_B,y_B}
  (\chi_{y_B} - \sum_y \omega_{y_B,y}\,\psi_y)\,].
\end{split}
\ee
Here $S$ and $S_B$ are the original and blocked actions, respectively,
$\omega_{x,x_B}$ describes the averaging process in a local neighborhood of
the coarse lattice point $x_B$, while $R$ is an arbitrary local operator whose
inverse is also local and diagonal in
Dirac space\footnote{The simplest example is 
$R^{-1}_{x_B,y_B}=\half \delta_{x_B,y_B}$}.
Take the fine lattice infinitely fine, so the field $\psi$ lives in the
continuum and $S$ is a chiral invariant action in the continuum. Since the
block transformation explicitly breaks chiral symmetry (it has a $\chibar
\dots \chi$ structure), so does the action $S_B$ on the lattice. 
The renormalization group transformation, however, does not change the long 
distance (in lattice units) behavior: $S_B$ should remember that the starting
action was chiral invariant. It is a few steps of algebra to dig out this
information~\cite{GW1982}. Writing 
\be
S_B= \sum_{x_B,y_B} \chibar_{x_B}\,D_{x_B,y_B}\,\chi_{y_B}
\ee
the Dirac operator is constrained by the relation
\be
\label{GW}
D \gamma^5 + \gamma^5 D = D \gamma^5 2R D \,.
\ee

The physical implications of the Ginsparg-Wilson relation in eq.~\eqref{GW}
which was obtained for free fermions
go far beyond of what the derivation above might suggest. Assume there
exists a local, gauge covariant solution of this equation in lattice QCD. The
first trivial observation is that the inverse of D satisfies
\be
\label{GW1}
D^{-1} \gamma^5 + \gamma^5 D^{-1} = \gamma^5 2R  \,.
\ee
The inverse of $D$, the quark propagator, from which the physical correlators
are constructed
is a non-local quantity. The equation
above shows, however, that its anticommutator with $\gamma^5$ is local,
i.e. has
an extension of $O(a)$ only. The violation of chiral symmetry is a contact
term. 
It has been shown in ref.~\cite{GW1982} also that a Dirac operator
satisfying  eq.~\eqref{GW} reproduces the $U(1)$ chiral anomaly. 

The following simple consideration strongly suggests that lattice QCD with a
Dirac 
operator satisfying eq.~\eqref{GW}  gives chiral invariant answers for all
the physical questions. Consider an arbitrary hadron correlator where the
operators are separated by physical distances. Write down the corresponding
Ward identity using the standard method in path integral formulation:
introduce new Grassmann integration variables which are related to the
original ones by an infinitesimal  $\gamma^5$ chiral rotation. The action is
not invariant and, according to eq.~\eqref{GW}, gives an insertion 
$\propto \chibar D \gamma^5 2R D \chi$ to the correlation function. The
fermion field $\chi$ ($\chibar$) will be paired with one of the antifermion
(fermion) fields of the hadron operators. The two propagators which bind the
breaking term to the hadron operators are canceled by the two $D$ operators in
the breaking term. What remains is $\gamma^5 2R_{x,y}$, where $x$ and $y$ are
the positions of the hadron operators. Since $R$ is local and the hadron
operators are separated at physical distances, the contribution
from the breaking term is zero. 

Eq.~\eqref{GW} is a highly non-trivial equation. The chance of finding a
local solution in the interactive case was rather dim.
The paper and its results were hardly
noticed and got completely forgotten\footnote{According to SLAC Spires the
paper was not cited at all over twelve years between 1986 and 1997 September
when the first solution of the GW equation in the presence of gauge fields was
identified.}. 

\section{Improvement programs}
On a lattice with finite (in physical units) lattice spacing $a$ the
predictions systematically deviate from the final $a \rightarrow 0$ continuum
values. To increase the resolution (decrease the cutoff effects) by a factor
of two in a given physical volume requires at least 16 times more computing
work and memory. Since the final continuum limit is universal but the cutoff
effects are not, the search for lattice actions with reduced cutoff effects
is an important issue since the beginnings. Although this question was raised
from the numerical side, the theory behind the improvement programs reveal
interesting aspects of quantum field theories and it is also part of the
developments around chiral symmetry. 

\subsection{Symanzik improvement}
To illustrate the basic idea consider a free massless scalar field with a
simple nearest neighbor action on the lattice
\be
S_0 = \sum_x  \half \partial_\mu \phi(x) \partial_\mu \phi(x)\,.
\ee
The pole of the propagator in Fourier space defines the energy as the function
of the momentum. For small momenta the energy has the form
\be
E^2 = {\bf p}^2-\frac{a^2}{12}\left(({\bf p}^2)^2 + \sum_{i=1}^3 p_i^4 \right)
   + O(a^4)\,,
\ee
where for better visual understanding the lattice unit $a$ is written out
explicitly. The leading cutoff effect is $O(a^2)$. It is easy to see that by
adding a dimension 6 ('irrelevant') term $cS_1$ to $S_0$, where
\be
S_1 = \sum_{x,\mu}  \half \partial_\mu \partial_\mu \phi(x) 
              \partial_\mu \partial_\mu \phi(x)\,
\ee  
and $c_1=1/12$ the $O(a^2)$ cutoff effect cancels. The action $S=S_0+c_1S_1$
is an $O(a^2)$ improved lattice action which approaches continuum limit
significantly faster than the original action $S_0$.

In the interacting case the situation is more involved.
As it is well known the coefficients of the higher dimensional operators 
enter the relation between the renormalized and the bare parameters. This
relation contains diverging powers and logs of the cutoff. Expressed in terms
of the renormalized parameters, however, the continuum predictions show no
sign of the presence of higher dimensional operators (universality). 
They have a role, however, in the improvement.

In spite of the technical complications, the final statement on improvement
is bold and simple: by adding to the original interactive action a
linear combination of the dimension 6 operators with properly chosen
coefficients, the $O(a^2)$ cutoff effects can be eliminated in all physical
quantities to all orders of perturbation theory. This has been shown by
Symanzik~\cite{Sym} for the $\phi^4$ scalar model and in the asymptotically
free $d=2$ non-linear $\sigma$-model. This is a highly non-trivial result
which has been obtained with the help of Callan-Symanzik type of
renormalization group equations and local effective lagrangians.

The improvement technique can be extended to gauge theories and
QCD~\cite{impr1}. During the last two decades large effort was invested to
calculate the improvement coefficients not only for the action but for
currents and other densities also. These coefficients can be calculated in
perturbation theory, or by simulations non-perturbatively. The Symanzik
improvement has an important role in controlling the cutoff
effects in different applications~\cite{Sharpe}.     

In QCD, if chiral symmetry is broken on the lattice,
the leading cutoff effects are $O(a)$ which makes improvement even more
important. $O(a)$ improvement can be achieved by adding a lattice version of
the dimension 5 operator $\psibar \sigma_{\mu\nu}F_{\mu\nu}\psi$ with a
tuned coefficient~\cite{impr1}. 
This term breaks chiral symmetry and has no place
in a chiral symmetric formulation. A chiral symmetric action is automatically
$O(a)$ improved making this possibility even more attractive. 

The $U(1)$ (non-singlet) chiral symmetry preserved by the 4-flavor staggered
fermions assures $O(a)$ improvement. Nevertheless, as painfully realized, this
action has large cutoff effects on lattices which are typically  simulated
today~\cite{stcut}. Although the
problem could be alleviated by canceling $O(a^2)$ flavor symmetry breaking
effects~\cite{PL}, the message remains.
Close to the continuum limit $O(a)$ ($O(a^2)$) cutoff effects dominate in
bosonic (fermionic) theories. Whether in today's simulations this is the case
depends on the theory, the space-time dimension, the form of the leading
action and on the form of the higher dimensional operators
chosen. Accidentally, 
just the $d=2$ asymptotically free $O(3)$ non-linear $\sigma$-model, which 
was used by Symanzik
to illustrate how to extend his method to theories with constraints, 
defies compliance with the Symanzik program in 
 numerical simulations~\cite{sigma}.

\subsection{Classically perfect actions}
There exist lattice actions which have no cutoff effects
whatsoever in the classical theory. Take the $d=2$ asymptotically free $O(3)$
non-linear $\sigma$-model mentioned above. One can put this theory on a
quadratic lattice with a local, classically perfect action. The corresponding
Euler-Lagrange equations have exact scale invariant instanton solutions, they
carry a topological charge $Q$ which is an integer, on any configuration the
value of the action $S \ge 4\pi |Q|$ and $S = 4\pi |Q|$ on solutions. If the
interaction is switched off the particles have the exact continuum dispersion
relation $E=|{\bf p}|$. The existence of such an action follows from Wilson's
renormalization group (RG) theory: the classically perfect action is the fixed
point (FP) of a RG transformation and is determined by {\it classical} saddle
point equations~\cite{HN}. 

The same is true for QCD in $d=4$~\cite{FPQCD,prospect}. Write the QCD 
action in the
form $\beta S_g+S_f$, where $S_g$ and $S_f$ denote the gauge and fermion
parts, $\beta \propto 1/g^2$ and consider the RG transformation
in QCD
\begin{multline}
\exp\bigl[-\bigl( \beta' S'_g(V)+S'_f(\bar{\chi},\chi,V)\bigr) \bigr]= \\
\int D\bar{\psi}D\psi DU \exp \bigl[ -\beta \bigl(S_g(U)+T_g(V,U)\bigr)+
\bigl(S_f(\bar{\psi},\psi,U)+T_f(\bar{\chi},\chi;\bar{\psi},\psi,U) \bigr)
\bigr]\, .
\label{qcdrgt}
\end{multline}
Here $T_g(V,U)$ defines the gauge invariant averaging process from the fine
$U$ to the coarse field $V$. The averaging function for the fermions 
$T_f(\bar{\chi},\chi;\bar{\psi},\psi,U)$ can be chosen in the form like in 
eq.~\eqref{GWblock} after introducing parallel transporters when adding
fermion fields in different lattice points in the averaging: $\omega
\rightarrow \omega(U)$ and similarly $R^{-1} \rightarrow R^{-1}(U)$. There 
is a considerable freedom here, the details are
not important. In an asymptotically free theory the FP is at 
$\beta \rightarrow \infty$ where the path integral above is reduced to
classical saddle point equations. For the gauge action this reads as
\begin{equation}
S^{{\rm FP}}_g(V) =\min_{\{U \}} \bigl[ S^{{\rm FP}}_g(U) + T_g(V,U) \bigr]\,,
\label{fpym}
\end{equation}
while for the fixed point Dirac operator in $S^{{\rm FP}}_f$ one obtains
\begin{equation}
D^{{\rm FP}}(V)^{-1}_{x_B,y_B} = R(V)_{x_B,y_B} +
\sum_{n,n'}\omega(U)_{x_B,x}D^{{\rm FP}}(U)^{-1}_{x,y}
\omega(U)^\dagger_{y,y_B} \, .
\label{fpD}
\end{equation}
Using the FP equation eq.~\eqref{fpym} for the gauge action $S^{{\rm FP}}_g$ 
one can derive results similar to that discussed above for the $O(3)$
$\sigma$-model on classical solutions, scale invariant instantons, etc.
at finite lattice resolution. 
The FP Dirac operator from  eq.~\eqref{fpD} has exact dispersion relation
in the free field limit and gives the correct continuum magnetic moment
(g-factor) independently of the lattice resolution.

The action $S^{{\rm FP}}_g$ is defined if for any configuration $V$ the value 
$S^{{\rm FP}}_g(V)$ (a real number) can be calculated. Take a configuration $V$ on
an $L^4$ lattice and assume that the blocking is from $L'=2^k L$ to
$L$. Considering any valid (gauge invariant, local) gauge action on the
r.h.s. of  eq.~\eqref{fpym} the minimum gives the value of $S^{{\rm FP}}_g(V)$
with an error which goes to zero as $\propto 4^{-k}$. If we insert
an approximate solution for the FP gauge action on the r.h.s., the error is
reduced further. The convergence of the solution of the FP Dirac operator in
eq.~\eqref{fpD} is $\propto 2^{-k}$.

Considering the action $\beta S^{{\rm FP}}_g + \psibar D^{{\rm FP}} \psi$ for 
{\it finite} $\beta$ we are off the renormalized trajectory (where the action
would be quantum perfect). It is expected nevertheless that a classically
perfect action will perform well in quantum simulations also. This is
demonstrated in different $d=2$ and 4 models including $SU(3)$ Yang-Mills
theory and in (quenched) QCD~\cite{fpsim}.  

In a numerical simulation, where we need the value the gauge action and the
$D^{{\rm FP}}v$ matrix-vector product frequently, there is no way to
calculate them using the FP equations above. One can, however, parametrize the
solutions in terms of gauge loops for $S_g$ and a certain number of fermion
offsets and paths for $D$, fit the coefficients to
eqs.~(\ref{fpym},\ref{fpD}) and obtain an approximate solution.

This improvement program is less systematic than Symanzik improvement. Very
close to the continuum limit the Symanzik program with non-perturbatively
determined coefficients is the best procedure. On today's lattices, however, 
the approximate FP actions in many cases perform better than Symanzik improved
actions~\cite{fpsim}. They are, however, also more expensive.  

As I shall discuss, the FP Dirac operator played a relevant role in the
quest for a chiral symmetric lattice regularizations also.

\section{Domain-wall and overlap fermions}
In 1992 Kaplan~\cite{Kaplan} suggested a new approach for lattice fermions
allowing them to move in a five-dimensional space. Setting up a
four-dimensional 
domain-wall some light modes become localized to the wall, the other,
non-localized modes remain heavy on the cutoff level. The idea is related to
earlier considerations in the continuum~\cite{DWcont}.

Consider a $d=5$ free Dirac operator in the continuum with an $x_5$ dependent 
mass term
\be
D=\gamma^\mu \partial^\mu+\gamma^5 \partial^5  - M(x_5)\,, 
\ee 
where $M(x_5 \rightarrow \pm \infty) = \pm m_0$ and forms a kink
between these values at $x_5=0$. The Dirac equation has a massless 
($ip_4=|{\bf p}|, {\bf p}=(p_1,p_2,p_3)$) left handed solution which is
localised to the $d=4$ kink at $x_5=0$. All the other modes live on the cutoff
level $m_0$. Kaplan realized that this construction can be taken over to the
lattice and the mechanism works in the presence of gauge fields
also. Actually, the five-dimensional model is not a genuine gauge-fermion
theory: the gauge fields know nothing about the fifth direction. It might be
better to consider this extra dimension as an internal 'flavor' space and the 
associated new fermion fields as regulators which are there to keep chiral
symmetry~\cite{NN}. In this approach, which is related to other
descriptions~\cite{FSRS} also, 
the effect of the massless chiral fermion can be
represented as an overlap of two fermionic states~\cite{NN}.

The domain-wall approach for vector theories like QCD has been streamlined
by Shamir~\cite{SSF}. The domain-wall was replaced by Dirichlet boundary
conditions. The left and right handed components of the Dirac fermion are
separated with one chirality bound exponentially on one wall, and the other on
the 
opposite wall. Using the notations in eqs.~(\ref{nnlat},\ref{gauder}) the
five-dimensional Dirac operator has the form
\be
D_{{\rm DW}} = \half \big[\gamma^5 (\partial^5 + \partial^{*5}) - 
\partial^5\partial^{*5}\big] + D_{{\rm W}} -m_0 \,,
\ee 
where $D_{{\rm W}}$ is the four-dimensional Dirac operator as defined in
eq.~\eqref{Dwilson}, and $0 < m_0 < 2$.\footnote{For $m_0<0$ there exist no
massless modes, for $m_0>2$ doublers appear.} 
The Dirichlet boundary conditions
set up in $x_5=0$ and $x_5=N$ read 
\be
\label{locferm}
P_R \chi(x_1,x_2,x_3,x_4,0) =0\,, \qquad P_L \chi(x_1,x_2,x_3,x_4,N) =0\,,
\ee
where $P_{R/L}=(1 \pm \gamma^5)/2$ and $\chi$ is the
five-dimensional fermion field. In the limit $N \rightarrow \infty$ the
resulting effective four-dimensional theory is chiral symmetric on the
lattice. I postpone the discussion on this most important point. 
Let me mention that, at the time of its discovery, it was not an issue
whether the domain-wall idea had anything
to do with the Ginsparg-Wilson relation.

This was a real breakthrough. Although $D_{{\rm DW}}$ has an extra dimension
(or 'flavor' space), its structure is quite similar to that of the 
four-dimensional Wilson action which raised the hope for 
a relatively simple generalization of the numerical
procedures. The computational cost is increased by a factor $O(N)$,
where $N$ should be large, but in problems where chiral symmetry plays an
essential role there might be no other way to proceed. Concerning the present
status of applications with domain-wall fermions I refer to the recent paper
in ref.~\cite{Blum}, while for an application with overlap fermions see 
ref.~\cite{Rebbi}. 

\section{The fixed-point Dirac operator in QCD solves the
  Ginsparg-Wilson equation}
It was noticed in 1997 that the fixed point Dirac operator in QCD 
satisfies the Ginsparg-Wilson relation \cite{prospect}. 
This conclusion
follows directly from the saddle-point equation eq.~\eqref{fpD}. Considering
the r.h.s. of eq.~\eqref{fpD} on a very fine lattice the fermion propagates 
over a very fine
gauge field configuration $U$. In this case the fermion propagator goes to its
continuum limit in this classical equation and the only term which is not 
anticommuting with $\gamma^5$ is $R(V)_{x_B,y_B}$. This leads to the equation
eq.~\eqref{GW1} (or eq.~\eqref{GW}) with a local $R$. In addition to
being classically perfect, the fixed point QCD action has this important
quantum property. The locality of the fixed-point action follows from general
arguments of Wilson renormalization group theory\footnote{Actually, one can
optimize the averaging function $\omega$ in eq.~\eqref{fpD} to make $D$ not
only local but decaying very fast~\cite{feri}.} 

In retrospect it is difficult to understand why this trivial observation was
not made earlier. The free fixed point Dirac operator was
explicitly known for different block transformations~\cite{UWPK}\footnote{An
example was already presented in the Appendix of ref.~\cite{GW1982} refering to
the unpublished PhD thesis of M.~Peskin. In 1982 the community was not yet
ripe to appreciate the Ginsparg-Wilson paper and pieces of its content were
independently rediscovered later.} Its physical content is just the
statement that $D_{{\rm FP}}$ is local, the energy-momentum dispersion relation
is exactly linear and the energy runs in $(0,\infty)$ (in the first Brillouin
zone) like in a chiral symmetric continuum formulation.

It was also illustrated in \cite{prospect} 
that the Ginsparg-Wilson relation implies Ward identities
from which the standard physical conclusions of chiral symmetry follow. 

\section{Index theorem on the lattice}
It is a simple exercise to show that the solutions of the Ginsparg-Wilson 
relation satisfy the index theorem on the lattice~\cite{HLN}: the zero modes
of $D$ are chiral and the associated index is a topological invariant which
represents the topological charge on the lattice.

For notational
simplicity consider the case $2R=1$ in eq.~\eqref{GW} and assume the
hermiticity property in Euclidean space $D^\dagger = \gamma^5 D \gamma^5$. It
is then easy to show that the spectrum $\{\lambda \}$ of $D$ lies on a unit
circle around the point $z=1$ in the complex plane with the following
properties. The real modes 
$\lambda=0$ and $\lambda=2$ are chiral (i.e. the corresponding eigenvectors are
also eigenvectors of $\gamma^5$ with $\pm 1$ eigenvalues). If $\phi_{\lambda}$
is an eigenvector with a complex eigenvalue $\lambda$ then  
$\phi_{\lambda^*}=\gamma^5 \phi_{\lambda}$ and they are orthogonal:
$(\phi_{\lambda},\gamma^5 \phi_{\lambda})=0$  

Define the density 
\be
q(x) = \half \tr(\gamma^5 D(x,x))\,,
\ee
where the trace is taken in color and Dirac space. Consider now
\begin{multline}
\sum_x q(x) =  \half \Tr(\gamma^5 D\big) = - \half \Tr\big[\gamma^5
(2-D)\big] \\
= -\half \sum_{\lambda} (2-\lambda)
( \phi_{\lambda}, \gamma^5 \phi_{\lambda})\,,
\end{multline}
where $\Tr$ denotes trace in all the indices and in the first step we
subtracted zero: $\Tr\gamma^5=0$. Using the orthogonality properties discussed
above only the $\lambda=0$ modes contribute giving
\be
\sum_x q(x)= n_{-}- n_{+}\,,
\ee
where $n_{-}$ ($n_{+}$) is the number of left (right) handed zero modes of the
Dirac operator. The index of $D$ is, therefore, a sum over $x$
of the pseudoscalar density
$q(x)$. It can be shown~\cite{FFtilde} that on smooth configurations $q(x)$ is
the continuum topological charge density
\be
q(x) =  \frac{1}{32 \pi^2} \epsilon_{\mu\nu\rho\sigma} 
\tr^c(F_{\mu\nu}F_{\rho\sigma}) + O(a^2)  \,,
\ee  
where the trace is taken in color space.
In the case of the fixed point Dirac operator one can show in addition that 
$\sum_x q(x)$ is the fixed-point topological charge $Q_{{\rm FP}}$ which takes
integer values on any gauge configuration, i.e. the topological charge from the
gauge and from the fermion sector is always consistent on the
lattice if the FP action is used. 

\section{Ginsparg-Wilson fermions from the domain-wall construction}
The simplicity of the index theorem demonstrated the power of the
Ginsparg-Wilson relation. It was obvious that this relation is relevant. 
In a short time a second solution, the 
overlap Dirac operator, was identified~\cite{Neu}. Unlike the fixed point Dirac
operator this solution has a simple explicit structure which allows
to represent it in a computer to machine precision. In addition, the overlap
operator binds the Ginsparg-Wilson approach to the domain-wall idea.

I present here this connection in a form as it occured after several
simplifying steps~\cite{domainsimp} and summarized in ref.~~\cite{HeJaLu}.
As eq.~\eqref{locferm} shows the $d=4$ fermion fields can be naturally
identified with
\be
\psi_x = P_L \chi(x,1) + P_R \chi(x,N)\,, \qquad
\psibar_x = \chibar(x,1)P_R + \chibar(x,N)P_L\,,
\ee
where $x \in Z_4$. The inverse of $\langle \psi_x \psibar_x\rangle$ 
defines an effective
four-dimensional Dirac operator $D^N_{x,y}$. In the $N \rightarrow \infty$
limit $D^N \rightarrow D$ can be written as
\be
\label{effdom}
aD = 1 - {\cal A}({\cal A}^\dagger {\cal A})^{-\frac{1}{2}}\,,
\ee
where
\be
{\cal A} = -a_5(D_{\rm W}-m_0) \left(1-\half a_5 (D_{\rm W}-m_0)\right)^{-1}
\ee   
and $a$ and $a_5$ are the lattice units in the four-dimensional space and in
the fifth direction, respectively. Taking the limit $a_5 \rightarrow 0$ one
obtains the overlap operator
\be
D_{\rm ov}= 1 - A( A^\dagger A)^{-\frac{1}{2}}\,,
\ee  
where $A= -(D_{\rm W} - m_0)$. Both $D_{\rm ov}$ and the Dirac operator $D$ in 
eq.~\eqref{effdom} satisfy the Ginsparg-Wilson relation with $2R=1$. This can
be shown easily by observing that both operators have the form $1-V$, where
$V$ is unitary.

A valid Dirac operator should be local, i.e. it should have an extension of
$O(a)$ (inverse cutoff). It has been shown~\cite{Horvath} that a solution of 
the
Ginsparg-Wilson equation can not be ultralocal\footnote{An ultralocal Dirac
operator has a finite number of fermionic offsets. The Wilson Dirac operator
$D_{\rm W}$, for example, is ultralocal having 9 offsets.}. Ultralocality is,
however, not necessary. Locality (and so, universality) requires that the
couplings in the action decay with the distance faster than any physical
signal. An exponential decay of the couplings like $\exp(-\kappa r/a)$, where
$\kappa$ is $O(1)$ defines a local operator. It has been shown~\cite{domloc}
that the overlap Dirac operator satisfies this bound if the theory on the 
lattice is close to the continuum limit.

\section{Exact chiral symmetry transformation}
As I discussed before, one can raise arguments that the Ginsparg-Wilson
relation implies the physical consequences of chiral symmetry. One can derive
Ward identities also and demonstrate that they lead to the same physical
predictions as those with exact chiral symmetry~\cite{PH1998}. 
This approach is, however,
cumbersome. In 1998 L\"uscher has found an exact symmetry transformation which
could be identified as the chiral symmetry transformation on the
lattice~\cite{L1998}. This was a theoretical breakthrough. It gave an elegant
technical tool to derive Ward identities, study anomalies and to handle the
problem of chiral gauge theories. Chiral symmetry is realized on the lattice in
an unusual, but beautiful way. 

Consider a vector gauge theory with $N_f$ flavors and take $2R=1$ in 
eq.~\eqref{GW} for simplicity.
If the Dirac operator $D$ satisfies the Ginsparg-Wilson relation then the
infinitesimal non-singlet transformation
\be
\label{nonsing}
\psi'=\psi +  i\epsilon T \gamma^5 \left(1-\half D\right)\psi\,,\qquad
\psibar'= \psibar + i\epsilon \psibar\left(1-\half D\right)\gamma^5 T
\ee
and the singlet transformation
\be
\label{sing}
\psi' = \psi + i\epsilon  \gamma^5 \left(1-\half D\right)\psi\,, \qquad
\psibar'= \psibar +  i\epsilon \psibar\left(1-\half D\right)\gamma^5 
\ee
leave the fermion action $\psibar D \psi$ invariant. In 
eq.~\eqref{nonsing} $T$ is an $SU(N_f)$ generator. It is trivial to demonstrate
this statement. Less trivial is the way this symmetry transformation
avoids the Nielsen-Ninomiya no-go theorem: the chiral transformation is not a
simple $\gamma^5$ rotation as the no-go theorem assumed.
The transformation depends on $D$
and so it depends on the gauge field over which the fermions propagate.  
In the formal
continuum limit the transformation goes over to the standard $\gamma^5$
rotation\footnote{In eqs.~(\ref{nonsing},\ref{sing}) 
$D$ is multiplied by the lattice
unit $a$ as it is obvious from dimensional analysis.}.

The action is invariant under the singlet transformation also, but the 
fermion integration measure is not~\cite{L1998}. 
A non-trivial Jacobi determinant is generated depending on the gauge
field
\be
\prod_x d\psibar'_x d\psi'_x = \left(1 +i \epsilon \Tr(\gamma^5 D)\right)
\prod_x d\psibar_x d\psi_x \,.
\ee
For a non-singlet transformation the Jacobi determinant is 1 due to 
$\tr(T) =0$.  
According to the index theorem on the lattice, a factor of 
$1+ i\epsilon 2 N_f Q_{\rm top}$ is produced by the measure for a singlet
transformation, where $Q_{\rm top}$ is the
topological charge of the gauge configuration as defined by the index of the
Dirac operator. Consider the transformation in eq.~\eqref{sing} as a change of
variable in the fermionic path integral. The expectation value 
$\langle{\cal O}\rangle$ of an arbitrary operator remains unchanged 
which leads to the correct anomalous Ward identity on the lattice
\be
\langle \delta {\cal O}\rangle + 
2 N_f \langle Q_{\rm top} {\cal O}\rangle = 0\,,
\ee
where $\delta {\cal O} =
\big[{\cal O}(\psibar',\psi')- {\cal O}(\psibar,\psi)\big]/ i \epsilon $ 
is the variation of the operator ${\cal O}$ 
under the transformation.

\section{Left and right handed fermions, the mass term and the $\theta$
parameter} 
In the formal continuum the massless fermion action of a vector theory falls
into left handed and right handed parts
\be
\label{decomp}
\psibar D \psi = \psibar_{L} D \psi_{L} + \psibar_{R} D \psi_{R}\,.
\ee
This is also possible on the lattice although in this case the projectors on
the $\psi$ field depend on the gauge configuration~\cite{Nar,ML99,feri}:
\be
\label{projectors}
\psi_{L}={\hat P}_L \psi\,,\qquad \psi_{R}={\hat P}_R\psi \,,\qquad
\psibar_{L}= \psibar P_R \,,\qquad \psibar_{R}=\psibar P_L \,,
\ee
where
\be
\label{gamma5hat}
{\hat P}_{R/L}=\half (1 \pm {\hat \gamma}_5)\,, \qquad
P_{R/L}=\half (1 \pm \gamma_5)\,.
\ee
Here ${\hat \gamma}_5$ depends on the gauge field through the Dirac operator
\be
{\hat \gamma}_5 = \gamma_5 ( 1 - D)
\ee
and satisfies $({\hat \gamma}_5)^2 = 1$ due to the Ginsparg-Wilson
relation (I consider $2R=1$ for simplicity). Using these definitions one
obtains the decomposition in  eq.~\eqref{decomp} on the lattice.

The relations above are striking in many ways. The decomposition 
$\psi(x) = \psi_L(x) + \psi_R(x)$ depends on the gauge field in the
neighborhood of $x$. Even for free fermions, the projectors depend on the
Dirac operator. Beyond that, there is an asymmetry between the fermions
and antifermions: the projectors on $\psibar$ are standard, they are identical
to those in the continuum. This asymmetry is deeply related to the fermion
number anomaly in chiral gauge theories~\cite{thooft}.

\noindent
The scalar and pseudo-scalar densities are introduced as usual
\begin{align}
S & = \psibar_L \psi_R + \psibar_R \psi_L = \psibar \left(1 - \half D\right)
 \psi\,, \\ 
P & = \psibar_L \psi_R - \psibar_R \psi_L = 
     \psibar \gamma_5 \left(1 - \half D\right) \psi\,. 
\end{align}
These densities behave correctly under chiral transformations. A
mass term in the action of a vector theory like QCD can be introduced as
\be
\label{fullaction}
S = S_g(U) + \int dx \left[\psibar D \psi + \psibar_L m \psi_R
+ \psibar_R m^\dagger \psi_L -i \theta q(x) \right]\,,
\ee
where a CP breaking $\theta$-term is added also and a sum over flavors is
assumed. For a real mass $m$ the massive Dirac operator can be written as
$(1-m/2)(D+m)$. The mass shifts the 
eigenvalue circle by $m$ and modifies the radius in such a way that the point
$\lambda=2$, where the ultraviolet eigenvalues are concentrated remains fixed.
Starting from eq.~\eqref{fullaction} the known Ward identities can be
derived, but this time in a controlled environment. The Ward identities are
valid at any
finite value of the cutoff (and so also in the continuum limit) as far as the
Dirac operator is a local solution of the Ginsparg-Wilson relation. Conserved
vector, axial vector and chiral currents can be defined~\cite{KiYa,HHHJN}. 
It became possible to clarify
theoretically opaque quantities like the topological 
susceptibility~\cite{Giu,Lu04} and to make
progress on the related Witten-Veneziano relation~\cite{Rossi}.
The Ginsparg-Wilson relation has also been used to construct actions with
improved chiral and scaling properties, like the chirally improved (CI)
action\cite{CI}.

\section{Chiral gauge theories}
Standard perturbatively defined regularizations have problems with
chiral gauge theories. Of course, in
principle one can give up gauge symmetry and introduce all the counter terms
needed to get gauge invariant renormalized predictions in
perturbation theory. This might not be the best way to proceed, not even
practically. In addition, our ultimate goal is to understand these theories
(like the standard model) beyond perturbation theory.

The decomposition eq.~\eqref{decomp} is the first step towards a gauge
invariant formulation of chiral gauge theories on the lattice. The projectors
in eq.~\eqref{projectors} are gauge field dependent, so the dividing surface
between the left and right handed fermions in the space of fermion degrees of
freedom is moving as the gauge configuration is changing. Even the number of
fermions with a certain chirality, say left handed, is changing. Since the
number of left handed antifermions is constant (their projectors are gauge
field independent), the total fermion number is changing as we are flying 
over the
gauge configuration space. Indeed, the difference between the fermion and
antifermion degrees of freedom is
\be
\label{fermionnumber}
\Tr{\hat P}_L - \Tr P_R = \half \Tr(\gamma_5 D) \,,
\ee  
which, according to the index theorem discussed before, takes different values
in the different topological sectors. On one hand, eq.~\eqref{fermionnumber}
shows that fermion number violation occurs naturally in this
formulation and this a welcome feature. On the other hand, having
different degrees of freedom in the different topological sectors might imply
difficulties to connect these sectors with each other in establishing the
theory.

In the last few years important results were obtained in this field. L\"uscher
has demonstrated that a U(1) chiral gauge theory with fermions in anomaly free
representation can be fully defined on the lattice so that gauge
symmetry is exactly preserved~\cite{U1}. The same could be established for a
general compact group in every order of perturbation theory\cite{general}. 
Beyond the $U(1)$ case it is not known whether and 
under which conditions these
theories are free of non-perturbative obstructions (anomalies). Such
obstructions might exist as has been shown in an $SU(2)$ chiral theory with a
single left handed fermion in the fundamental representation~\cite{wittensu2}.
This was also demonstrated on the lattice~\cite{Bar}. A different approach
which is pursued on the lattice since a long time is to give up explicit
gauge symmetry. I refer here to a recent paper~\cite{SHGO} and to the
references therein. 

\section{Closing remarks}
The Metropolis algorithm also celebrated its fiftieth
anniversary recently~\cite{LosAla}.
Nobody thought fifty years ago
that Yang-Mills theory, this beautiful theoretical construction, will have so
much to do with a stochastic algorithm as it is the case since almost three
decades.  
These simulations delivered a lot of non-perturbative results on the glueball
spectrum, static potential and the related fluctuating flux tube and
thermodynamics. The results are quantitative and some of them are quite
precise. The numerical results can be connected to analytic predictions in
some corners of the theory (for a highly non-trivial example I refer to the
contribution of van Baal in this volume). 

QCD is, of course, more difficult and much more exciting. If the doubts
concerning the validity of 
the running staggered fermion simulations~\cite{qcdsolved} are
positively clarified we shall certainly see a large number of interesting
non-perturbative QCD predictions during the next few years. I think, this will
happen in any case soon. 

{\bf Acknowledgements}

I thank Anna Hasenfratz and  Ferenc Niedermayer for the discussions and 
the enjoyable collaboration over the years. This work was supported by the
Schweizerischer Nationalfonds.


\eject



\begin{thebibliography}{99}

\bibitem{YM}
C.~N.~Yang and R.~L.~Mills, Phys.~Rev.~{\bf 96}, 191 (1954).

\bibitem{FP}
L.~D.~Fadeev and V.~N.~Popov, Phys.~Lett.~{\bf B25}, 29 (1967).

\bibitem{HV}
G.~'t~Hooft, Nucl.~Phys.~{\bf B33}, 173 (1971);
Nucl.~Phys.~{\bf B35}, 167 (1971);
G.~'t~Hooft and M.~Veltman, Nucl.~Phys.~{\bf B44}, 189 (1972).

\bibitem{PGW}
H.~D.~Politzer, Phys.~Rev.~Lett. {\bf 30}, 1346 (1973);
D.~J.~Gross and F.~Wilczek, Phys.~Rev.~Lett. {\bf 30}, 1343 (1973).

\bibitem{qcd}
H.~Fritzsch, M.~Gell-Mann and H.~Leutwyler, Phys.~Lett.~{\bf B47}, 365 (1973);
S.~Weinberg, Phys.~Rev.~Lett. {\bf 31}, 494 (1973).

\bibitem{Wilson1974}  
K.~G.~Wilson, Phys.~Rev.~{\bf D10}, 2445 (1974).

\bibitem{BDI74_75}
R.~Balian, J.~M.~Drouffe and C.~Itzykson, Phys.~Rev.~D10, 3376 (1974); 
 Phys.~Rev.~{\bf D11}, 2098 (1975);  Phys.~Rev.~{\bf D11}, 2104 (1975).

\bibitem{KS1975}
J.~Kogut and L.~Susskind, Phys.~Rev.~{\bf D11}, 395 (1975).

\bibitem{CJR1979}
M.~Creutz, L.~Jacobs and C.~Rebbi, Phys.~Rev.~Lett. {\bf 42}, 1390 (1979);
Phys.~Rev.~{\bf D20}, 1915 (1979.

\bibitem{Creutz1980}
M.~Creutz, Phys.~Rev.~Lett. {\bf 45}, 313 (1980); 
Phys.~Rev.~{\bf D21}, 2308 (1980).

\bibitem{tube2004}
M.~L\"uscher and P.~Weisz, JHEP~{\bf 0207}, 049 (2002) [arXiv:hep-lat/0207003];
K.~J.~Juge, J.~Kuti and C.~Morningstar, 
Phys.~Rev.~Lett. {\bf 90}, 161601 (2003) [arXiv:hep-lat/0207004].

\bibitem{DeG}
T.~DeGrand, Int.~J.~Mod.~Phys. {\bf A19}, 1337 (2004) [arXiv:hep-lat/0312241].

\bibitem{feri}
F.~Niedermayer, Nucl.~Phys.~Proc.~Suppl. {\bf 73},105 (1999)
[arXiv:hep-lat/9810026].

\bibitem{summaries}
M.~Creutz, Rev.~Mod.~Phys. {\bf 73}, 119 (2001) [arXiv:hep-lat/0007032];
S.~Chandrasekharan and U.-J.~Wiese, arXiv:hep-lat/0405024.

\bibitem{giulu}
M.~L\"uscher, Nucl.~Phys.~Proc.~Suppl. {\bf 83} 34 (2000)
[arXiv:hep-lat/9909150]; 
L.~Giusti, Nucl.~Phys.~Proc.~Suppl. {\bf 119}, 
149 (2003)[arXiv:hep-lat/0211009].

\bibitem{Reisz}
T.~Reisz,  Nucl.~Phys.~{\bf B318}, 417 (1989).

\bibitem{KaSm}
L.~H.~Karsten and J.~Smit, Nucl.~Phys.~{\bf B192}, 100 (1981).

\bibitem{Adler}
S.~L.~Adler, Phys.~Rev.~{\bf 177}, 2426 (1969).

\bibitem{BJ}
J.~S.~Bell and R.~Jackiw, Nuovo Cim. {\bf 60A}, 47 (1969). 

\bibitem{NN1981}
N.~B.~Nielsen and M.~Ninomiya, Nucl.~Phys.~{\bf B185}, 20 (1981).

\bibitem{WilsonErice}
K.~G.~Wilson, in {\it New Phenomena in Subnuclear Physics}, ed. A.~Zichichi,
(Plenum Press, New York) Part A, 69 (1977).

\bibitem{staggered}
L.~Susskind,  Phys.~Rev.~{\bf D16}, 3031 (1977).

\bibitem{qcdsolved}
C.~T.~H.~Davies et al. Phys.~Rev.~Lett. {\bf 92}, 022001 (2004)
[arXiv:hep-lat/0304004].

\bibitem{GW1982}
P.~H.~Ginsparg and K.~G.~Wilson,  Phys.~Rev.~{\bf D25}, 2549 (1982).

\bibitem{Sym}
K.~Symanzik, Nucl.~Phys.~{\bf B226}, 187, 205 (1983).

\bibitem{impr1}
M.~L\"uscher and P.~Weisz,  Nucl.~Phys.~{\bf B240}, 349 (1984);
M.~L\"uscher and P.~Weisz,  Commun.~Math.~Phys.~{\bf 97}, 59 (1985);
R.~Wohlert and B.~Sheikholeslami, Nucl.~Phys.~{\bf B259}, 572 (1985).

\bibitem{Sharpe}
S.~R.~Sharpe, in {\it Vancouver 1998, High energy physics}, vol. 1* 171,
[arXiv:hep-lat/9811006]. 

\bibitem{stcut}
S.~Aoki,  Nucl.~Phys.~Procs.~Suppl.{\bf 94}, 3 (2001) [arXiv:hep-lat/0011074].

\bibitem{PL}
G.~P.~Lepage, Phys.~Rev.~{\bf D59}, 074502 (1999) [arXiv:hep-lat/9809157].

\bibitem{sigma}
P.~Hasenfratz and F.~Niedermayer, Nucl.~Phys.~{\bf B596},~481~(2001) 
[arXiv:hep-lat/0006021];
\newline
M.~Hasenbusch et al., Nucl.~Phys.~Procs.~Suppl.{\bf 106}, 911 (2002) 
[arXiv:hep-lat/0110202].

\bibitem{HN}
P.~Hasenfratz and F.~Niedermayer, Nucl.~Phys.~{\bf B414}, 785 (1994) 
[arXiv:hep-lat/9308004].

\bibitem{FPQCD}
T.~DeGrand, A.~Hasenfratz, P.~Hasenfratz and F.~Niedermayer, 
Nucl.~Phys.~{\bf B454}, 587 (1995) [arXiv:hep-lat/9506030];
W.~Bietenholz and U.-J.~Wiese, Nucl.~Phys.~{\bf B464}, 319 (1996)
[arXiv:hep-lat/9510025];
T.~DeGrand, A.~Hasenfratz, D.~Zhu, Nucl.~Phys.~{\bf B475}, 321 (1996)
[arXiv:hep-lat/9603015].

\bibitem{prospect}
P.~Hasenfratz,  Nucl.~Phys.~Procs.~Suppl.{\bf 63A-C}, 53 (1998)
[arXiv:hep-lat/9709110]. 

\bibitem{fpsim}
F.~Niedermayer, P.~R\"ufenacht and U.~Wenger, Nucl.~Phys.~{\bf B597}, 
413 (2001) [arXiv:hep-lat/0007007];
C.~Gattringer et al., BGR-Collaboration, Nucl.~Phys.~{\bf B677},
3 (2004) [arXiv:hep-lat/0307013].

\bibitem{Kaplan}
D.~B.~Kaplan, Phys.~Lett. {\bf B288}, 342 (1992) [arXiv:hep-lat/9206013].

\bibitem{DWcont}
V.~A.~Rubakov and M.~E.~Shaposhnikov, Phys.~Lett. {\bf B125}, 136 (1983);
C.~G.~Callan and J.~A.~Harvey, Nucl.~Phys.~{\bf B250}, 427 (1985). 

\bibitem{NN}
R.~Narayanan and H.~Neuberger, Phys.~Lett. {\bf B302}, 62 (1993) 
[arXiv:hep-lat/9212019] ;
Phys.~Rev.~Lett. {\bf 71}, 3251 (1993) [arXiv:hep-lat/9308011];
Nucl.~Phys.~{\bf B412}, 574 (1994)[arXiv:hep-lat/9307006];
Nucl.~Phys.~{\bf B443}, 305 (1995) [arXiv:hep-lat/9411108].

\bibitem{FSRS}
S.~Frolov and A.~Slavnov, Nucl.~Phys.~{\bf B411}, 647 (1994)
[arXiv:hep-lat/9303004]; 
S.~Randjbar-Daemi and J.~Strathdee, Nucl.~Phys.~{\bf B443}, 386 (1995)
[arXiv:hep-lat/9501027].

\bibitem{SSF}
Y.~Shamir, Nucl.~Phys.~{\bf B406}, 90 (1993) [arXiv:hep-lat/9303005];
V.~Furman and Y.~Shamir, Nucl.~Phys.~{\bf B439}, 54 (1995)
[arXiv:hep-lat/9505004].

\bibitem{Blum}
Y.~Aoki et al.,
Phys.~Rev.~{\bf D69}, 074504 (2004) [arXiv:hep-lat/0211023].

\bibitem{Rebbi}
N.~Garron, L.~Giusti, Ch.~Hoelbling, L.~Lellouch and C.~Rebbi,
Phys.~Rev.~Lett. {\bf 92}, 042001 (2004) [arXiv:hep-lat/0306295].

\bibitem{UWPK}
U.-J.~Wiese, Phys.~Lett. {\bf B315}, 417 (1993) [arXiv:hep-lat/9306003];
W.~Bietenholz and U.-J.~Wiese,
Nucl.~Phys.~Procs.~Suppl.{\bf 34}, 516 (1994) [arXiv:hep-lat/9311016];
P.~Kunszt, Nucl.~Phys.~{\bf B516}, 402 (1998) [arXiv:hep-lat/9706019]. 

\bibitem{HLN}
P.~Hasenfratz, V.~Laliena and F.~Niedermayer, 
Phys.~Lett. {\bf B427}, 125 (1998) [arXiv:hep-lat/9801021].

\bibitem{FFtilde}
K.~Fujikawa, Nucl.~Phys.~{\bf B546}, 480 (1999) [arXiv:hep-lat/9811235];
H.~Suzuki, Prog.~Theor.~Phys {\bf 102}, 141 (1999) 
[arXiv:hep-lat/9812019];
K.~Fujikawa, Int.~J.~Math.~Phys. {\bf B16}, 1931 (2002) 
[arXiv:hep-lat/0205024];
D.~H.~Adams, Annals Phys. {\bf 296}, 131 (2002) [arXiv:hep-lat/9812003].
 
\bibitem{Neu}
H.~Neuberger, Phys.~Lett. {\bf B427}, 353 (1998) [arXiv:hep-lat/9801031]. 

\bibitem{domainsimp}
H.~Neuberger, Phys.~Rev.~{\bf D57}, 5417 (1998) [arXiv:hep-lat/9710089];
A.~Borici, Phys.~Lett. {\bf B453}, 46 (1999) [arXiv:hep-lat/9810064];  
Y.~Kikukawa and T.~Noguchi, arXiv:hep-lat/9902022.

\bibitem{HeJaLu}
P.~Hern\'andez, K.~Jansen and M.~L\"uscher, arXiv:hep-lat/0007015.

\bibitem{Horvath}
I.~Horvath, Phys.~Rev.~Lett. {\bf 81}, 4063 (1998) [arXiv:hep-lat/9808002]. 

\bibitem{domloc}
P.~Hern\'andez, K.~Jansen and M.~L\"uscher, 
Nucl.~Phys.~{\bf B552}, 363 (1999) [arXiv:hep-lat/9808010];
M.~Golterman and Y.~Shamir,Phys.~Rev.~{\bf D68}, 074501 (2003) 
[arXiv:hep-lat/0306002]. 

\bibitem{PH1998}
P.~Hasenfratz, Nucl.~Phys.~{\bf B525}, 401 (1998) [arXiv:hep-lat/9802007].

\bibitem{L1998}
M.~L\"uscher, Phys.~Lett. {\bf B428}, 342 (1998) [arXiv:hep-lat/9802011].

\bibitem{Nar}
R.~Narayanan, Phys.~Rev.~{\bf D58}, 097501 (1998) [arXiv:hep-lat/9802018]. 

\bibitem{ML99}
M.~L\"uscher, Nucl.~Phys.~{\bf B538}, 515 (1999) [arXiv:hep-lat/9808021]. 

\bibitem{thooft}
G.~'t~Hooft, Phys.~Rev.~{\bf D14}, 3432 (1976). 

\bibitem{KiYa}
Y.~Kikukawa and A.~Yamada, Nucl.~Phys.~{\bf B547}, 413 (1999) 
[arXiv:hep-lat/9810024]. 

\bibitem{HHHJN}
P.~Hasenfratz, S.~Hauswirth, K.~Holland T.~J\"org and F.~Niedermayer,
Nucl.~Phys.~{\bf B643}, 280 (2002) [arXiv:hep-lat/0205010].

\bibitem{CI}
C.~Gattringer, Phys.~Rev.~{\bf D 63}, 114501 (2001),
[arXiv:hep-lat/0003005];
C.~Gattringer, I.~Hip and C.~B.~Lang, Nucl.~Phys.~{\bf B597}, 451 (2001),
[arXiv:hep-lat/0007042].

\bibitem{Giu}
L.~Giusti, G.~C.~Rossi, and M.~Testa,
Phys.~Lett. {\bf B587}, 157 (2004) [arXiv:hep-lat/0402027].

\bibitem{Lu04}
M.~L\"uscher, arXiv:hep-lat/0404034.

\bibitem{Rossi}
L.~Giusti, G.~C.~Rossi, M.~Testa and G.~Veneziano,
Nucl.~Phys.~{\bf B628}, 234 (2002) [arXiv:hep-lat/0108009].

\bibitem{U1}
M.~L\"uscher, Nucl.~Phys.~{\bf B549}, 295 (1999) [arXiv:hep-lat/9811032];
Nucl.~Phys.~{\bf B568}, 162 (2000) [arXiv:hep-lat/9904009].

\bibitem{general}
H.~Suzuki, Progr.~Theor.~Phys. {\bf 101}, 1147 (1999)[arXiv:hep-lat/9901012]; 
Nucl.~Phys.~{\bf B585}, 471 (2000) [arXiv:hep-lat/0002009];
M.~L\"uscher, JHEP~{\bf 06} 028 (2000) [arXiv:hep-lat/0006014].

\bibitem{wittensu2}
E.~Witten, Phys.~Lett. {\bf B117}, 324 (1982).

\bibitem{Bar}
O.~B\"ar and I.~Campos, 
Nucl.~Phys.~Procs.~Suppl.{\bf 83}, 594 (2000) [arXiv:hep-lat/0001025].

\bibitem{SHGO}
M.~Golterman and Y.~Shamir, arXiv:hep-lat/0404011.

\bibitem{LosAla}
AIP Conference Proceedings 690, (2003),
{\it The Monte Carlo Method in the Physical Sciences: Celebrating the 50th
Anniversary of the Metropolis Algorithm}. 


\end{thebibliography}
\end{document}